\documentclass[traditabstract]{aa}
\usepackage{natbib}
\usepackage[pdftex]{graphicx}
\usepackage{amsmath}
\usepackage{color}
\usepackage{hyperref}
\makeatletter
\usepackage{subfigure}

\def\alfven{Alfv\'{e}n}
\def\tapp{\theta_{\rm app}}
\def\lesssim{\mathrel{\hbox{\rlap{\hbox{\lower4pt\hbox{$\sim$}}}\hbox{$<$}}}}
\def\gtrsim{\mathrel{\hbox{\rlap{\hbox{\lower4pt\hbox{$\sim$}}}\hbox{$>$}}}}
\begin{document}
\title{Causal connection in parsec-scale relativistic jets: results from the MOJAVE VLBI survey}
\titlerunning{Causal connection in relativistic jets}

\author{E. Clausen-Brown\inst{\ref{inst1}}\thanks{clausenbrown@mpifr.de} \and T. Savolainen\inst{\ref{inst1}}\and A.\,B. Pushkarev \inst{\ref{inst2}}\inst{,\ref{inst3}}\inst{,\ref{inst1}} \and Y.\,Y. Kovalev \inst{\ref{inst4}}\inst{,\ref{inst1}}\and J.\,A. Zensus\inst{\ref{inst1}}}
\institute{Max-Planck-Institut f\"{u}r Radioastronomie, Auf dem H\"{u}gel 69, 53121 Bonn, Germany\label{inst1}
\and Crimean Astrophysical Observatory, 98409 Nauchny, Crimea, Ukraine\label{inst2}
\and Pulkovo Astronomical Observatory, Pulkovskoe Chaussee 65/1, 196140 St. Petersburg, Russia\label{inst3}
\and Astro Space Center of Lebedev Physical Institute, Profsoyuznaya 84/32, 117997 Moscow, Russia\label{inst4}}




\abstract{We report that active galactic nucleus (AGN) jets are causally connected on parsec scales, based on 15 GHz Very Long Baseline Array (VLBA) data from a sample of 133 AGN jets.  This result is achieved through a new method for measuring the product of the jet Lorentz factor and the intrinsic opening angle $\Gamma\theta_j$ from measured apparent opening angles in flux density limited samples of AGN jets. The $\Gamma\theta_j$ parameter is important for jet physics because it is related to the jet-frame sidewise expansion speed and causal connection between the jet edges and its symmetry axis.  Most importantly, the standard model of jet production requires that the jet be causally connected with its symmetry axis, implying that $\Gamma\theta_j\la1$.   When we apply our method to the MOJAVE flux density limited sample of radio loud objects, we find $\Gamma\theta_j\approx 0.2$, implying that AGN jets are causally connected.  We also find evidence that AGN jets viewed very close to the line of sight effectively have smaller intrinsic opening angles compared with jets viewed more off-axis, which is consistent with Doppler beaming and a fast inner spine/slow outer sheath velocity field. Notably, gamma-ray burst (GRB) jets have a typical $\Gamma\theta_j$ that is two orders of magnitude higher, suggesting that different physical mechanisms are at work in GRB jets compared to AGN jets.  A useful application of our result is that a jet's beaming parameters can be derived.  Assuming $\Gamma\theta_j$ is approximately constant in the AGN jet population, an individual jet's Doppler factor and Lorentz factor (and therefore also its viewing angle) can be determined using two observable quantities: apparent jet opening angle and the apparent speed of jet components.}


\maketitle

\section{Introduction}
A wide variety of processes in relativistic outflows are sensitive to causal connection, or the ability of a disturbance at the edge of an axisymmetric flow to communicate with the symmetry axis.\footnote{Another type of causal connection that we do not discuss in this paper is the ability of a disturbance to communicate upstream with the jet's central engine.  For super-fast magnetosonic jets, these disturbances cannot propagate back to the central engine.}  In the standard magnetic model of relativistic jet production \citep{Beskin2010}, the global jet structure determines the nature of bulk acceleration, implying that the jet must be causally connected for such acceleration to take place \citep[e.g.,][]{Tchekhovskoy2009,Komissarov2009}.  Causal connection is determined by the half opening angle of an axisymmetric flow, $\theta_j$, and the flow Lorentz factor, $\Gamma$, through their product $\Gamma\theta_j$, where $\Gamma\theta_j\la1$ implies that the jet is causally connected.  Other important aspects of relativistic jet physics that depend on $\Gamma\theta_j$ include jet stability \citep{Narayan2009}, magnetic reconnection \citep{Giannios2013}, recollimation shock energy dissipation \citep{Nalewajko2009}, and recollimation shock structure \citep{Kohler2012a}.

Relativistic outflows have a wide range of values of $\Gamma\theta_j$.  On the one hand, pulsar wind nebulae contain uncollimated (equitorial) outflows that are inferred to reach very high bulk Lorentz factors of $\Gamma\sim 10^6$ prior to the termination shock \citep{Kennel1984}, implying that these outflows have values of $\Gamma\theta_j\sim 10^6$ and are not causally connected.  On the other hand, narrow relativistic outflows (i.e., jets) associated with X-ray binaries (XRBs), gamma-ray bursts (GRBs), and active galactic nucleus (AGN) typically have much lower values of $\Gamma\theta_j$; GRB light curve analyses lead to typical inferred values of $\Gamma\theta_j$ of $10-30$ \citep{Panaitescu2002}, while the narrowness and moderate apparent speeds in XRB jets support the assumption that $\Gamma\theta_j \la1$ \citep[e.g.,][]{Miller-Jones2006}.  Thus, while the central engines of the abovementioned objects are similar in that they involve compact magnetized spinning objects, their different values of $\Gamma\theta_j$ suggest that different physical processes are at work.

There have been two past measurements of the characteristic value of $\Gamma\theta_j$ for AGN jets.  Using 7\,mm Very Long Baseline Array (VLBA) data from 15 different AGN jets, \cite{Jorstad2005} measured $\Gamma\theta_j$ by assuming that the observed pattern speed of moving jet components corresponds to the jet bulk flow speed, and that the component variability times are equal to the jet frame light crossing times of the resolved components.  From these assumptions they determined the component's Lorentz factor and jet half-opening angle, and found an anti-correlation between the derived values of $\theta_j$ and $\Gamma$, with $\Gamma\theta_j=0.17$.  With a larger sample of 56 AGN jets from 15 GHz VLBA data, \cite{Pushkarev2009} performed the same analysis, except they used jet parameters from \cite{Hovatta2009}, who derived these values from variability time, maximum flux density of flares, and equipartition derived brightness temperature arguments.  The \cite{Pushkarev2009} analysis found a similar anti-correlation with $\Gamma\theta_j=0.13$.

Motivated by the above theoretical concerns, we construct a very different method of inferring $\Gamma\theta_j$ by using apparent opening angles obtained from a flux density limited sample of AGN jets.  We construct a theoretical probability density function for apparent opening angles in a flux density limited sample, which we derive in Sect.\,\ref{pdf}, and which has $\Gamma\theta_j$ as a free parameter to be fixed by finding the best fit to an empirical distribution of apparent opening angles.  Our data consist of the stacked images from 135 AGN jets that make up the MOJAVE-I sample, a 15 GHz flux density limited survey conducted by the VLBA of radio sources in the northern sky with flux densities above 1.5 Jy, and above 2 Jy for sources with $-20<$ dec $<0$ \citep{Lister2009}.  The stacked images and opening angles are also discussed and analyzed in \cite{Pushkarev2012a}.  After analyzing the data in Sect.\,\ref{data}, we discuss in Sect.\,\ref{discussion} the physical significance of the parameter $\Gamma\theta_j$ for AGN in the context of jet instabilities, GRB jet acceleration vs. AGN jet acceleration, and jet parameter estimation.  We conclude in Sect.\,\ref{conclusion}.

\section{Statistical model of \mbox{\boldmath{$\tapp$}}}
\label{pdf}
Here we model a given jet's value of $\tapp$ as a random variable that is drawn from the probability density function (PDF) $P(\tapp)$.  That is, $P(\tapp)d\tapp$ represents the probability that a given jet in a flux density limited sample will have an observed apparent half opening angle between $\tapp$ and $\tapp+d\tapp$.  First, however, we motivate our model by estimating $\Gamma\theta_j$ for blazars, and discuss the effect that velocity shear may have on jet appearance.

Blazars are oriented such that the angle between the jet symmetry axis and the line sight, $\theta_{ob}$, is $x/\Gamma$, where $x\approx 0.5$ on average in flux density limited samples \citep{Vermeulen1994} such as MOJAVE.  This value for $x$ implies an upper limit on $\Gamma\theta_j$ of 
\begin{align}
\Gamma\theta_j\la0.5. \notag
\end{align}
This is based on the simple argument that most MOJAVE jets are not observed ``down the pipe," ($=\theta_{ob}<\theta_j$), since a down-the-pipe AGN jet would not display jet-like morphology.  In fact, most MOJAVE sources do display a jet-like morphology, which implies that typically $\theta_{ob}>\theta_j$  \citep{Clausen-Brown2011}, and therefore that $0.5/\Gamma>\theta_j$ according to the typical value of $\theta_{ob}$ for flux density limited samples.  Also, an estimate of $\Gamma\theta_j$ can be made,
\begin{align}
\Gamma\theta_j&\sim0.10 \left(\frac{\langle \tapp \rangle}{0.2\text{ rad}}\right)\left(\frac{x}{1/2}\right),
\label{blazar}
\end{align}
where $\langle \tapp \rangle\approx0.2$ rad is the average apparent opening angle in the MOJAVE-I sample that we use in this work.  From geometrical considerations, as long as all the relevant angles are small, $\theta_j=\theta_{ob}\tapp$, thus if $\langle \tapp \rangle$ is used for $\tapp$ and $0.5/\Gamma$ for $\theta_{ob}$, then we obtain Eq.\,(\ref{blazar}).  An interesting feature of the above estimate is that it does not significantly depend on the actual value of $\Gamma$, which is useful since jets possess a wide range of $\Gamma$ values \citep{Lister2009b}.  In Sect.\,\ref{prob_section} we will make a more rigorous analysis of the likely value of $\Gamma\theta_j$ for blazars in which we will also find that this estimate is mostly independent of blazar values of $\Gamma$.

A possibility we explore below is that there is a viewing angle effect related to Doppler beaming and velocity shear that affects the appearance of blazars.  Velocity shear is included in a variety of AGN jet models, including parsec-scale models \citep{Swain1998,Attridge1999,Tavecchio2008,Perucho2012},  kiloparsec-scale models \citep[e.g.][]{Owen1989,Swain1998,Perlman1999,Laing2004}, and more general jet models \citep{Aloy2000,Chiaberge2000,McKinney2006}.  We assume velocity shear affects very long baseline interferometry (VLBI) measurements of a jet's apparent opening angle $\tapp$.  While all jets may have the same value of $\Gamma\theta_j$, for jets viewed with very small viewing angles the emission might originate from a fast (beamed) narrow spine that is a fraction of the true jet opening angle $\theta_j$, while for jets with larger viewing angles the emission from a slow outer sheath with half-opening angle $\theta_j$ may be more detectable.  This effect is described in Sect.\,\ref{shear}.


\subsection{Derivation of $P(\tapp)$}
\label{prob_section}
To test the viability of the simplest case scenario, we assume that $\Gamma\theta_j$ is constant for all relativistic jets, and that these jets are conical and non-accelerating.  In general, however, jets are not conical and the jet flow is either accelerating or decelerating, although in some jet models $\Gamma\theta_j$ nevertheless remains constant \citep[e.g.,][]{Zakamska2008}.  Individual observed jet component acceleration is consistent with only very small changes in Lorentz factor, $\dot{\Gamma}/\Gamma\sim 10^{-3}$ \citep{Homan2009}, although an individual jet does posses a range of component speeds \citep{Lister2009b,Lister2013}.  Thus, our assumption of conical non-accelerating jets clearly introduces uncertainty to our model.

As we show below, $\Gamma\theta_j$ is a free parameter of $P(\tapp)$, and thus will be determined in the fit to the empirical distribution of $\tapp$.  To derive $P(\tapp)$, we first derive the PDF for viewing angles, $P(\theta_{ob})$.  If the sample of AGN jets is unbiased with respect to orientation, $P(\theta_{ob})=\sin\theta_{ob}$.  However, because it is flux density limited, it will take the form
\begin{equation}
P(\theta_{ob})=\mbox{Doppler bias factor} \times \sin\theta_{ob}.  \notag
\end{equation}
This additional factor takes into account that more sources are directed at the observer in a flux density limited sample because Doppler beamed jets are detectable at greater distances than unbeamed ones.  \cite{Cohen1989} and \cite{Vermeulen1994} computed this term and found that it depends on the bulk Lorentz factor distribution in a flux density limited sample, the integral source count index, and the beaming index of the jet.  The beaming index is defined from the relation $F=\delta^nF'$, where $n$ is the beaming index, $\delta$ is the Doppler factor, $F$ is the observed flux density density, and $F'$ is the intrinsic flux density.  The observed integral source count index is defined in the expression $N(>F)\propto F^{-q}$, representing the number of sources $N$ observed with a flux density above $F$, which is a power law in $F$ with source count index $q$.  Including the Doppler bias factor in the viewing angle PDF gives
 \begin{equation}
 P(\theta_{ob},\Gamma) =A \left(1-\beta\cos\theta_{ob}\right)^{-a-1}\sin\theta_{ob}P(\Gamma),
 \label{VC}
 \end{equation}
where $a=nq-1$, $P(\Gamma)$ is the PDF for jet bulk Lorentz factor, and $A$ is the normalization constant.  An important assumption made in calculating the Doppler bias term is that the log-log slope of $N(>F')$ vs. $F'$ and $N(>F)$ vs. $F$ are the same, which \cite{Vermeulen1994} justify based on previous studies of AGN jet luminosity functions \citep{Urry1984,Urry1991}.  The MOJAVE selection criteria were designed so that source inclusion in the sample is based on beamed emission only \citep{Lister2005}.  These MOJAVE sources are typically dominated by core flux density, which usually has a flat spectrum \citep{Kovalev2005,Pushkarev2012b}.  Thus, the beaming index $n$ is most likely $\sim 2$ for steady jets \citep{Lind1985}, while the integral source count index is approximately $1.5$, indicating that the fiducial value for the $a$-parameter should be approximately \citep{Vermeulen1994}
\begin{align}
a_{\rm fiducial}=2. \notag
\end{align}
We note that if the MOJAVE sources were typically dominated by optically thin flux density, which typically has a spectral index of $\alpha\sim0.7$ ($F_{\nu}\propto \nu^{-\alpha}$), then $a\approx 3$.

The opening angle distribution may now be derived from $P(\theta_{ob},\Gamma)$ by a change of variables from $\theta_{ob}$ to $\tapp$ and marginalizing over $\Gamma$,
\begin{equation}
P(\tapp)=\int{d\Gamma P\left(\theta_{ob}(\tapp,\Gamma),\Gamma\right)\left| \frac{\partial \theta_{ob}}{\quad\partial \tapp} \right|},
\label{ptapp}
\end{equation}
where $\theta_{ob}$ and $\partial \theta_{ob}/\partial \tapp$ are functions of $\tapp$ and $\Gamma$, and can be determined by assuming a particular jet geometry that we take to be conical here.  These relationships are often derived by treating conical jets as triangles projected onto the plane of the sky, implying that $\tan{\theta_{\rm app}}=R_j/\ell'=R_j/(\ell\sin\theta_{ob})$, where $\theta_{ob}$ is the jet viewing angle, $R_j$ is the jet radius, $\ell$ is the jet length, and $\ell'$ is the jet length projected onto the sky.  If we assume $\theta_j\approx R_j/\ell$, then
\begin{equation}
\tan\theta_j=\tan\tapp\sin\theta_{ob}.
\label{tri}
\end{equation}
For cases where both $\tapp$ and $\theta_j$ are $ \ll 1$, this reduces to a commonly used relation for jets, $\theta_j=\tapp \sin\theta_{ob}$ \citep{Jorstad2005,Pushkarev2009}.  Eq.\,(\ref{tri}) also implies a maximum apparent half-opening angle of $\pi/2$.  We assume that jets viewed down-the-pipe where $\theta_{ob}<\theta_j$ are rare, since such jets would have $\tapp>\pi/2$.  This dearth of down-the-pipe jets, sometimes used to justify the cylindrical approximation in jet models \citep{Clausen-Brown2011}, also indicates that typically $\Gamma\theta_j<1$.  If the typical viewing angle of a jet is $\theta_{ob}=0.5/\Gamma$ \citep{Vermeulen1994}, and most MOJAVE jets exhibit a jet-like morphology (i.e., $\tapp<\pi/2$) such that $\theta_{ob}>\theta_j$, then $\Gamma\theta_j<0.5$. Here, because many jets have large apparent opening angles, but are often viewed with small observing angles and small intrinsic opening angles, we most often use the approximation that 
\begin{align}
\theta_{ob}\approx\frac{\rho}{\Gamma\tan\tapp}.
\label{approx_theta_ob}
\end{align}  
This approximation is mostly appropriate for the blazar dominated MOJAVE sample; below in Sect.\,\ref{shear} we show that this approximation is useful for categorizing jets by their apparent opening angles.

An apparent weakness in our model is that $P(\Gamma)$ is not well constrained.  This is not the case, however, since $P(\tapp)$ is insensitive to $P(\Gamma)$, which we demonstrate here.  In a flux density limited VLBI sample, jets with small viewing angles will dominate, so we assume $\sin\theta_{ob}\approx \theta_{ob}$, and approximate (\ref{VC}) as
\begin{equation}
P(\theta_{ob},\Gamma)=A (2\Gamma^2)^{a+1}(1+\Gamma^2\theta_{ob}^2)^{-a-1}\theta_{ob} P(\Gamma).
\label{approx}
\end{equation}
For simplicity, we now evaluate Eq.\,(\ref{ptapp}) in light of the geometry implied by Eq.\,(\ref{approx_theta_ob}), and obtain
\begin{align}
P(\tapp)&=A\left(1+\frac{\rho^2}{\tan^2{\tapp}}\right)^{-a-1}\frac{\cos{\tapp}}{\sin^3{\tapp}} \notag\\
&\times\left[2^{a+1}\rho^2\int{\Gamma^{2a}P(\Gamma)d\Gamma}\right] \notag\\
&= A'\left(1+\frac{\rho^2}{\tan^2{\tapp}}\right)^{-a-1}\frac{\cos{\tapp}}{\sin^3{\tapp}},
\label{approx2}
\end{align}
where $\rho=\Gamma\theta_j$, and we have absorbed the term in brackets into the new normalization, $A'$.  Thus, it is apparent from Eq.\,(\ref{approx2}) that $P(\tapp)$ does not depend significantly on the form of $P(\Gamma)$.  Eq.\,(\ref{approx2}) is an accurate approximation of $P(\tapp)$ as long as $\rho\ll1$, which is a valid assumption as shown by our best value of $\rho\approx 0.2$ discussed below.

 \subsection{Velocity shear and Doppler beaming}
 \label{shear}
As suggested by the very approximate estimate made above in which radio galaxies appear to have larger $\theta_j$ than blazars, velocity shear and Doppler beaming may affect the distribution of $\tapp$.  Unfortunately, modeling the effect of velocity of shear on jet appearance is sensitive to a variety of unknown details regarding the jet structure such as how the density of non-thermal electrons scales with jet radius and the particular functional form of the velocity shear.  Thus, in an effort to capture only the most basic effect velocity shear has on jet appearance, we develop a minimalist model.  

\begin{figure}
\centering
\includegraphics[width=3.5in, trim=0cm 6cm 0cm 0cm, clip=true]{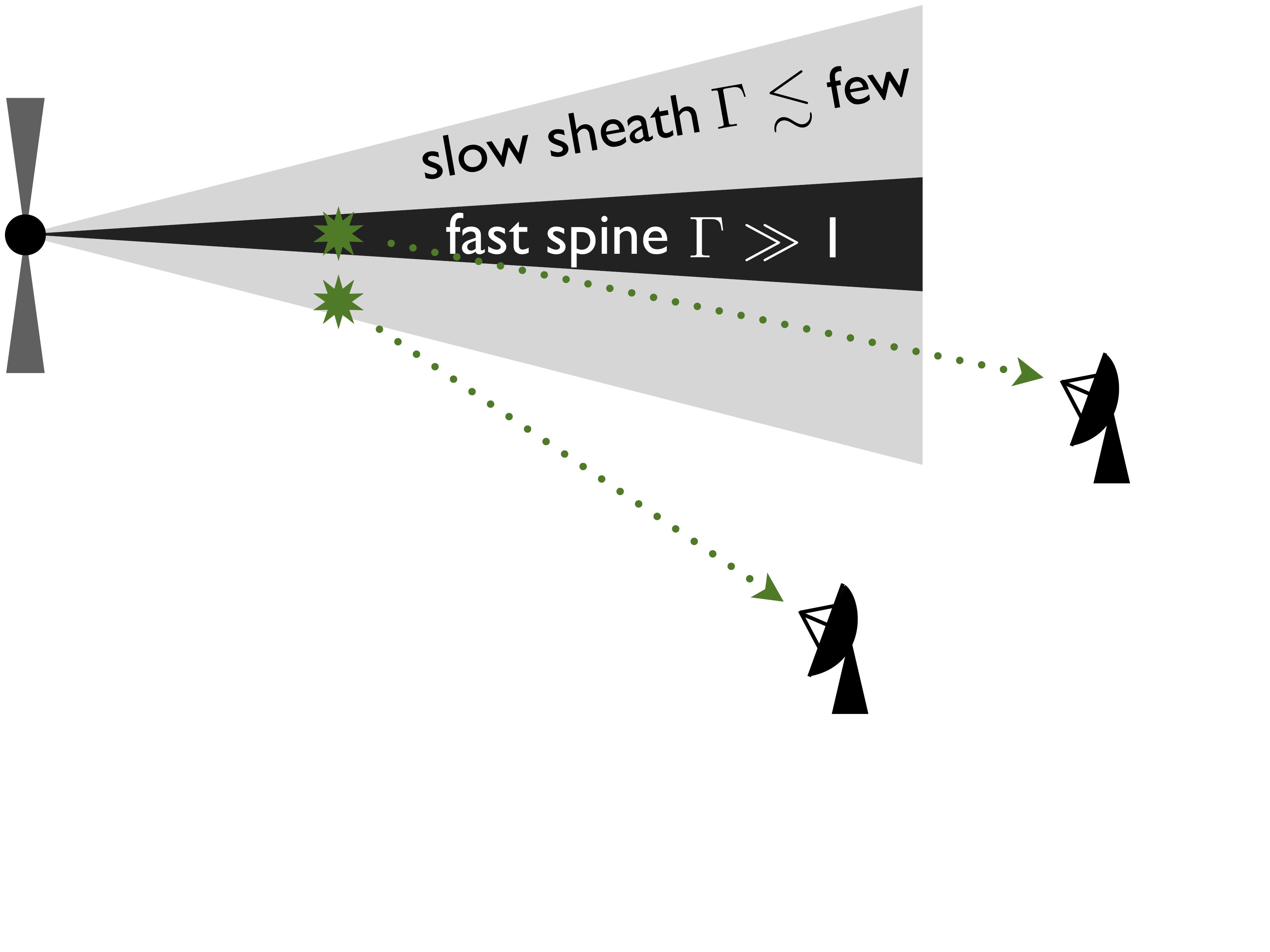}
\caption{Meridional slice of a jet illustrating two cases: (i) jets aligned close to the line of sight where emission is dominated by the fast inner spine, and (ii) more misaligned jets where the emission from outer slower layers contributes as well.}  
\label{cartoon}
\end{figure}

\begin{figure}
\centering
\includegraphics[width=3.3in]{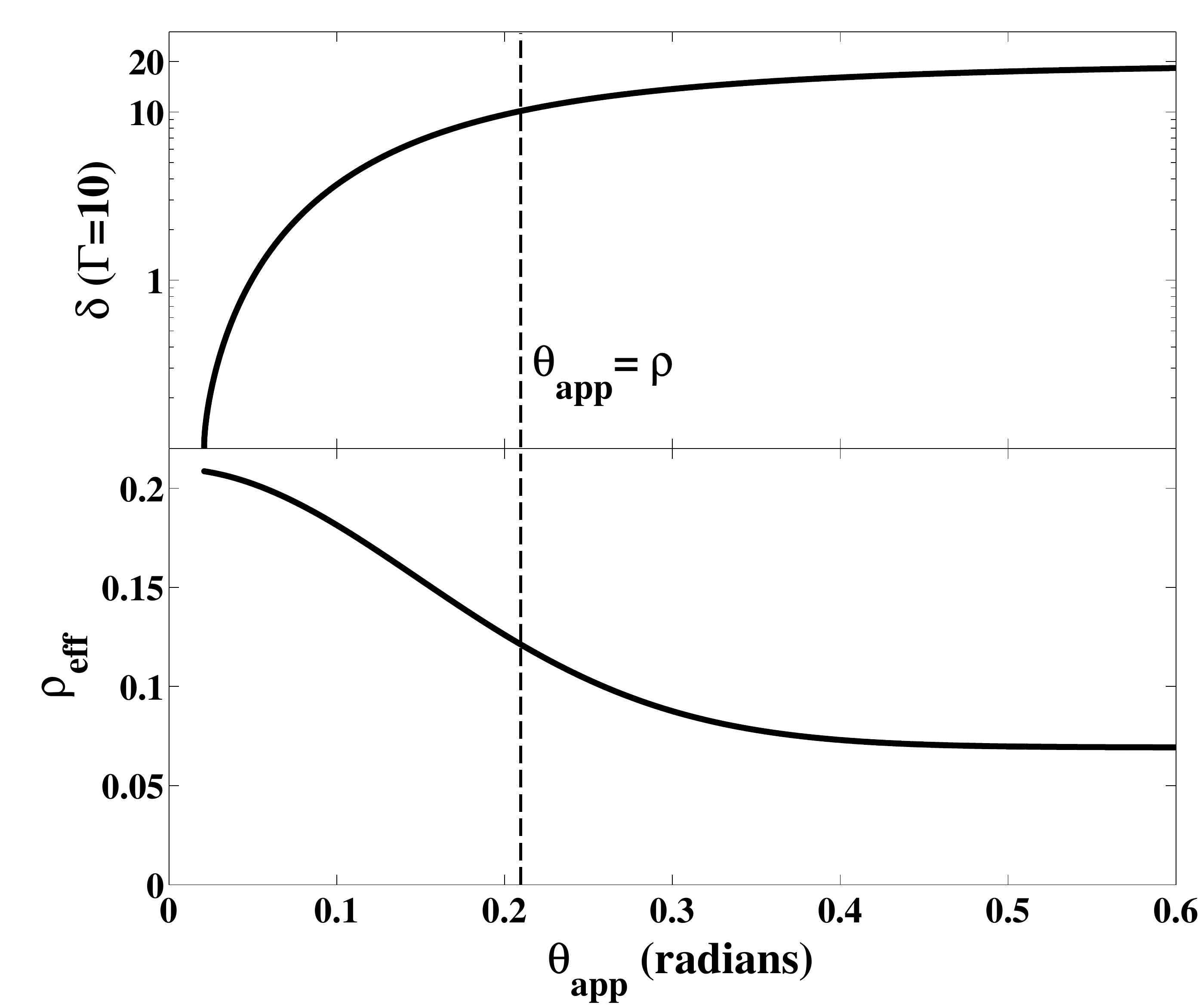}
\caption{Plots of Doppler factor and $\rho_{\rm eff}$ as a function of a jet's apparent half-opening angle $\tapp$.  The semi-logarithmic Doppler factor plot in the upper panel is for a jet with $\Gamma=10$ and $\rho=0.21$, and uses Eq.\,(\protect\ref{tri}) to convert $\tapp$ to $\theta_{ob}$.  The lower panel plot shows $\rho_{\rm eff}$ as a function of $\tapp$ from Eq.\,(\protect\ref{rho_eff}) using $\rho=0.21$ and $f_{\rm sh}=0.33$, the best fit values found in Sect.\,\ref{analysis}.}  
\label{shear_fig}
\end{figure}

We assume the velocity field of a jet consists of an ultra-relativistic inner spine and a surrounding shear layer that is mildly relativistic (see Fig. \ref{shear_fig}).  The relativistic spine, which dominates the core emission, is what primarily determines whether a jet is included in a flux density limited sample like the blazar dominated MOJAVE sample.  We note, however, that this assumption may sometimes be violated since  a few MOJAVE sources such as M87 may have significant sheath emission \citep[e.g.,][]{Kovalev2007}.  The optically thin jet downstream from the core is where apparent opening angles are measured, and where the degree to which the shear layer is observable is important.  For jets aligned close to the line of sight, the jets will be more dominated by the fast spine where $\Gamma\gg 1$ and the slower outer layers will remain unobserved, while more misaligned the jets will have a slower outer sheath of $\Gamma_{\rm shear}\la$ few that is more likely visible.  See Fig.\,\ref{cartoon} for an illustration of this effect.  

Now, jets can be divided into two categories based on whether a jet's viewing angle $\theta_{ob}$ is less than or greater than $1/\Gamma$, where $\Gamma$ is the value of the Lorentz factor in the fast inner spine.  This categorization can be mapped onto $\tapp$ by using Eq.\,(\ref{approx_theta_ob}), resulting in
\begin{align}
\tapp &>\arctan(\rho) \Longleftrightarrow \theta_{ob}<1/\Gamma \notag \\
\tapp&<\arctan(\rho) \Longleftrightarrow \theta_{ob}>1/\Gamma.
\end{align}
This categorization is useful since the critical angle $1/\Gamma$ defines when beaming is important.  When $\theta_{ob}>1/\Gamma$, then Earth is outside of the inner jet's beaming cone, thus the jet's slower outer layers are more likely to be visible, since the fast inner spine's beaming is less dominant.  The hypothesis that highly beamed jets ($\theta_{ob}<1/\Gamma$) and not highly beamed jets ($\theta_{ob}>1/\Gamma$) can be separated by their observed $\tapp$ has some observational support, which we discuss in Sect.\,\ref{dopplerbeaming}. 

The simplest way to model the effect that velocity shear and beaming have on jet appearence is to postulate that jets that have $\theta_{ob}<1/\Gamma$ have an effective jet opening angle $\theta_{j, \rm eff}=\rho_{\rm eff}/\Gamma$, where $\theta_{j, \rm eff}$ is some fraction of the true jet opening angle such that $\theta_{j,\rm eff}=f_{\rm sh}\theta_j$, or equivalenty, $\rho_{\rm eff}=f_{\rm sh}\rho$.  Thus,  
\begin{align}
\theta_{j, \rm eff} &\rightarrow \theta_j && \text{if } \tapp \ll \arctan(\rho), \notag \\
\theta_{j, \rm eff} &\rightarrow f_{\rm sh}\theta_j && \text{if } \tapp \gg \arctan(\rho), \notag
\end{align}
where $f_{\rm sh}$ is a free parameter.  We note that the inner spine Doppler factor of a jet with $\tapp\sim \theta_j$ (i.e., a radio galaxy) is $\delta \sim 1/\Gamma$, a jet with $\tapp=\rho$ has $\delta\sim \Gamma$, and a jet with $\tapp\sim 1$ has $\delta\sim 2\Gamma$.  In other words, the most drastic change in $\delta$ occurs in a narrow range of $\tapp$, for $0<\tapp<\rho$, while $\delta$ only changes by a factor of 2 in the large range $\rho<\tapp\la 1$.  To illustrate this point, we plot the Doppler factor as a function of $\tapp$ in Fig. \ref{shear_fig} .  Thus, the effect of velocity shear on jet appearance should be strongest for $\tapp=0$ to $\rho$.  To reproduce this behavior, we choose the following arbitrary function,
\begin{align}
\frac{\rho_{\rm eff}}{\rho}&=(1-f_{\rm sh})\exp{\left(-\left(\frac{\tapp}{\arctan(\rho)}\right)^2\right)}+f_{\rm sh}.
\label{rho_eff}
\end{align}
We plot this function in Fig.\,\ref{shear_fig}.  This equation can easily be inserted into our theoretical PDF described in Eq.\,(\ref{ptapp}), which can then be evaluated numerically, where $\rho$ and $f_{\rm sh}$ are free parameters to be found in the fit.  If this model is correct, then the best fit value of $f_{\rm sh}$ should be less than unity.  In the case of no shear, then $f_{sh}=1$ and $\rho_{\rm eff}=\rho$.

\section{Data analysis and results}
\label{data}
\subsection{Apparent opening angles}
\label{data_analysis}
The apparent opening angles used here are derived from stacked images of 133 sources from the MOJAVE-I catalogue of 135 sources.\footnote{\url{http://www.physics.purdue.edu/MOJAVE/allsources.html}}  For two sources opening angles could not be derived.  To produce a stacked image of a given source, all single-epoch maps were aligned by their VLBI core components, and then averaged together.  The resulting opening angle data originates from the analysis in \cite{Pushkarev2012a}, who derived $\tapp$ by taking the median value of 
\begin{align}
\tapp=\arctan\left(\frac{\sqrt{d^2-b_{\phi}^2}}{2r}\right),
\end{align}
where ``$d$ is the full width half maximum (FWHM) of the Gaussian transverse profile, $r$ is the distance to the core along the jet axis, $b_{\phi}$ is the beam size along the position angle $\phi$ of the jet-cut, and the quantity $(d^2-b_{\phi}^2)^{1/2}$ is the deconvolved FWHM transverse size of the jet" \citep[for more details, see][]{Pushkarev2012a}.  Note that in this work we use half opening angles, while \cite{Pushkarev2012a} used full opening angles, which merely differ by a factor of 2.

\subsection{Best fits and goodness of fit}
\label{analysis}
We now compare the opening angle data to Eq.\,(\ref{ptapp}), where $P(\Gamma)\propto \Gamma^{-1.5}$ with $\Gamma_{\rm min}=2$ and $\Gamma_{\rm max}=50$.  We note, however, that Eq.\,(\ref{ptapp}) is insensitive to the form of $P(\Gamma)$ as we demonstrated in Eq.\,(\ref{approx2}).

We find the best fits using maximum likelihood estimation (MLE) by minimizing the negative log-likelihood function
\begin{align}
h(\mathbf{X},\mathbf{m})=-2\displaystyle \sum_{i=1}^{N} \ln P(X_i, \mathbf{m}),
\label{logL}
\end{align}
where our data set is $\boldsymbol{X}=(\theta_{{\rm app},1},...,\theta_{{\rm app},N})$, $P(X_i,\mathbf{m})$ represents $P(\tapp)$ evaluated at $\tapp=X_i$, and $\mathbf{m}$ is a vector representing the free parameters of the distribution $P(\tapp)$.  As discussed below, we fit our data set of $N=133$ for six different cases in which the distribution's free parameters ranges from three, $\boldsymbol{m}=(\rho, f_{\rm sh}, a)$, to only one, $m=\rho$.  When $f_{\rm sh}$ is not a free parameter it is fixed at $1$, and when $a$ is not free it is fixed at either 2 or 3, as specified below.  In all of these cases, we obtain the best fit parameters $\boldsymbol{\hat{m}}$ by numerically minimizing $h$ (Eq.\,\ref{logL}).

To correctly model the fitting error and assess the goodness of fit, we used the Kolmogorov-Smirnov (KS) statistic $L_N$ in conjunction with the nonparametric bootstrap as described in \cite{Feigelson2012}.  Recall that the KS statistic gives a measure of the distance between the data and the model by finding the maximum distance between the empirical cumulative distribution function $F_N(\tapp)$ and theoretical cumulative distribution function $F(\tapp,\mathbf{\hat{m}})$, i.e.,
\begin{align}
\frac{L_N}{\sqrt{N}}=\sup_{\tapp} \left|F_N(\tapp)-F(\tapp,\mathbf{\hat{m}})\right|.
\label{KS}
\end{align}
Here, for each bootstrap realization, we generate the simulated data $\mathbf{X}^\ast$ via sampling with replacement, find the best fit parameters $\mathbf{\hat{m}}^\ast$ for the simulated data $\mathbf{X}^\ast$ using the MLE procedure described above, and then calculate the KS statistic $L_N^\ast$ from $\mathbf{X}^\ast$ and $\mathbf{\hat{m}}^\ast$ by using Eq.\,(\ref{KS}) with an additional bias correction factor taken into account \citep[see Eq.\,3.48 of][]{Feigelson2012}.

After iterating the bootstrap $B=2000$ times, we obtain confidence intervals around $\mathbf{\hat{m}}$ by analyzing the distribution of simulated best-fit parameters $\mathbf{\hat{m}}^\ast$ and directly compute the 68\% and $95\%$ confidence intervals and error contours.  The resulting distribution of the statistic $L_N^\ast$ can be used for model selection by finding the probability $p$ that a value of $L_N$ or greater is observed, assuming that $\mathbf{X}$ is drawn from $P(\tapp,\mathbf{\hat{m}})$.  This is done by defining $k$ as the number of $L_N^\ast$ values that fulfill the criterion $L_N^\ast\geq L_N$, and then computing the p-value as $p=k/B$.  Thus, for a significance level of $\alpha=0.05$, models with p-values of 0.05 and above are favored by the data (i.e. they cannot be rejected).  As discussed below, we consider six different cases, thus we perform different bootstrap simulations for each case.

\begin{figure}
\centering
\includegraphics[width=3.5in]{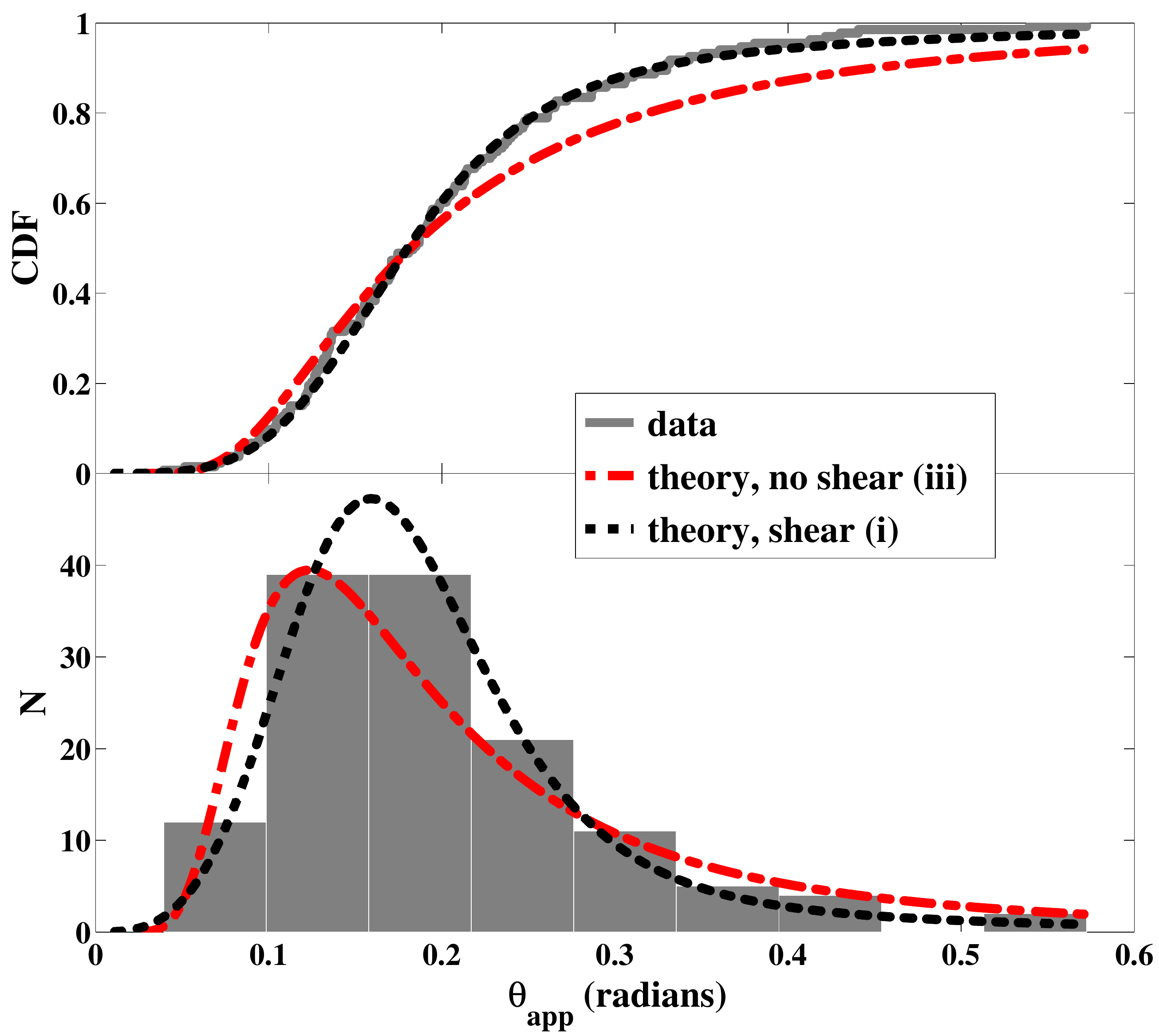}
\caption{Example best fits for two cases are shown in the form of cumulative distribution functions (CDF, upper panel) and probability density functions (PDF, lower panel).  The two cases shown, ``no shear (iii)" (dash-dotted line) and ``shear (i)" (dotted line), both have the same number of free parameters (two), but the shear (i) model clearly fits the data better.  The data is represented as a solid line (upper panel) or histogram (lower panel).}  
\label{datafit}
\end{figure}

\begin{figure}
\centering
\includegraphics[width=3.5in]{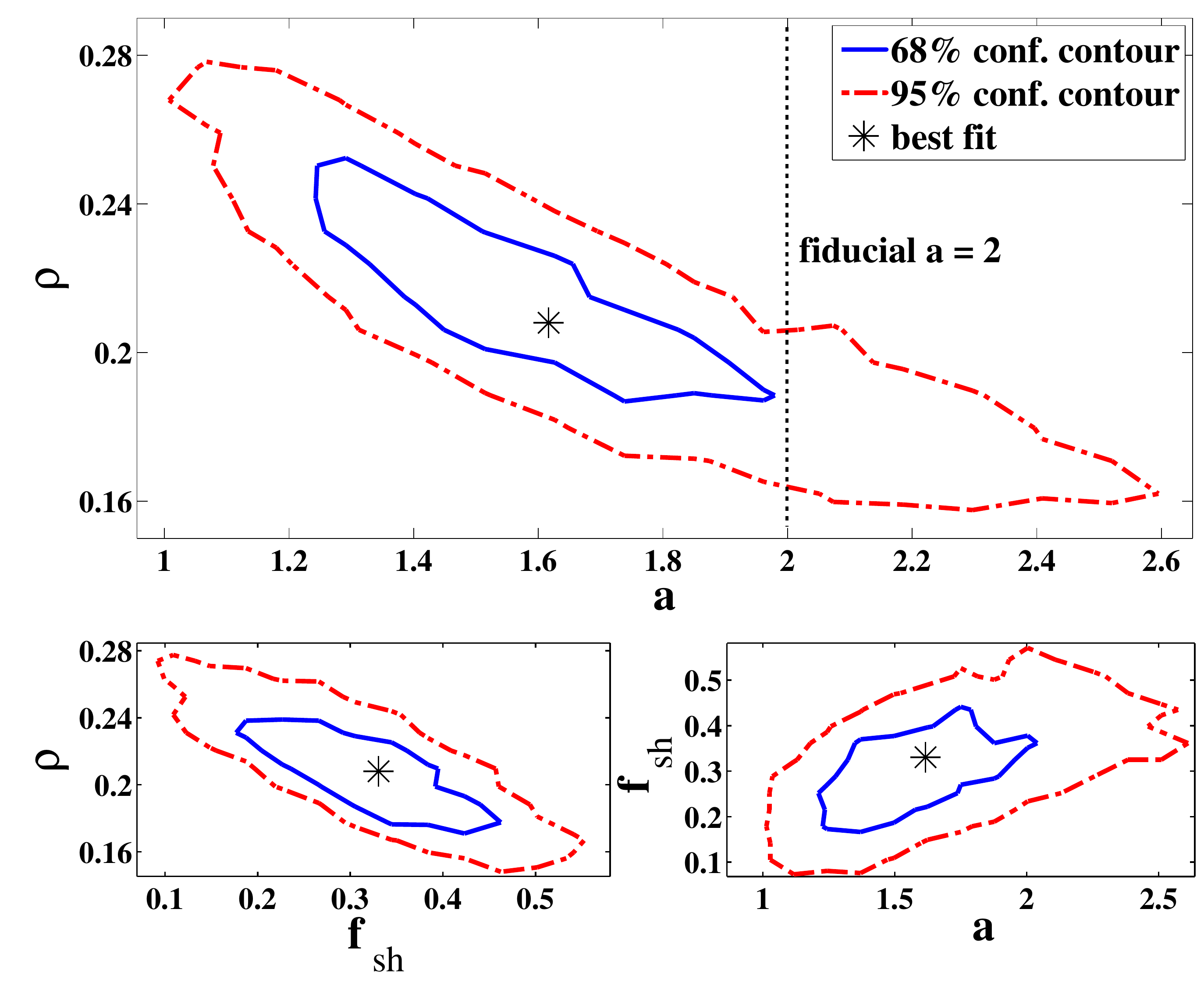}
\caption{68\% and 95\% confidence contours from Monte Carlo error analysis for the best-fit parameters $\rho$, $a$, and $f_{\rm sh}$, from the case of shear (iii).  The central star shows the MLE of the parameters.  We emphasize the $a$ vs. $\rho$ plot since there is a fiducial value for $a$ ($=2$, vertical dotted line), and $\rho$ has constraints set on it by \protect\cite{Jorstad2005} and \protect\cite{Pushkarev2009}.}  
\label{error_analysis}
\end{figure}

\begin{table} 
 \caption{Parameter values (both best fit and assigned) for six different cases along with 68\% confidence intervals where relevant.  }
 \label{symbols}
 \begin{tabular}{@{}lcccccc}
  \hline
  \rule{0pt}{2ex} 
   model& $\rho$ ($=\Gamma\theta_j$)& $a$ 
        & $f_{\rm sh}$ & p-val \rule{0pt}{2ex}\\
       
  \hline 
 \rule{0pt}{2.5ex}no\,shear\,(i)&$0.12\pm 0.005$ & $\equiv 2$ & $\equiv 1$ & $\la 0.001$  \\
  no\,shear\,(ii)& $0.095\pm 0.004$ & $\equiv 3$ & $\equiv 1$ & $\la 0.001$  \\
  no\,shear\,(iii)& $0.063\pm 0.016$ & $6.3^{+5.1}_{-2.2}$ & $\equiv 1$ & 0.004 \\
        shear\,(i) & $0.19\pm0.012$ & $\equiv 2$&$0.39 \pm 0.08$ & 0.064 \\
        shear\,(ii) & $0.15\pm 0.013$ & $\equiv 3$ & $0.5 \pm 0.1$ & 0.28 \\
        shear\,(iii) & $0.21\pm 0.03$ & $1.6^{+0.5}_{-0.3}$ & $0.33 \pm 0.1$ & 0.12  \\
        
  \hline
   \rule{0pt}{3ex}
 \end{tabular}
 \textbf{Notes.} The $\equiv$ sign preceding a parameter value means this value was pre-assigned, not found in the best fit. 
 \label{bestfits}
\end{table}

We now apply the above analysis to the data for two different types of models:
 
\emph{No shear model:} Three cases are considered for our model with no shear, i.e., $f_{\rm sh}=1$, depending how the parameter $a$ is treated: (i) $a$ is set to 2, (ii) $a=3$, and (iii) $a$ is a free parameter found in the best fit.  As it turns out, in case (iii) the best-fit value of $a$, 6.3, is much higher than the expected fiducial value of 2, and the distribution of best-fit values of $a$ in bootstrap simulations routinely ranges much higher (several tens).  In addition, the best values of $\rho$ in the bootstrap simulations is tightly correlated with $a$, and also ranges widely,  suggesting that $a$ and $\rho$ are highly degenerate. 

\emph{Velocity shear model:} Here we perform the same analysis and consider the same three cases as above, but with $f_{\rm sh}$ as a free parameter.  Thus, $\rho$ and $f_{\rm sh}$ are free parameters and we consider three different cases regarding $a$: (i) $a=2$, (ii) $a=3$, and (iii) $a$ as a free parameter to be found in the best fit.  

Table\,\ref{bestfits} gives a summary of our best-fit results, and Fig.\,\ref{datafit} shows a comparison between the data and two different best-fit models, each with the same number of free parameters (two).  The models that include relativistic velocity shear are clearly favored by the data, as they all have p-values above 0.05.  When shear is included and all the parameters are varied (the shear\,(iii) model), the best fit produces a reasonable value of $a=1.6^{+0.5}_{-0.3}$ (and $\rho=0.21 \pm0.03$ and $f_{\rm sh}=0.33 \pm 0.1$), which is close to the expected value of $a=2$ from Doppler beaming models and integral source counts of radio-selected AGN.  Encouragingly, the best-fit values for all of the models is $\rho=0.1-0.2$, which is consistent with the values of $0.17$ and $0.13$ reported in \cite{Jorstad2005} and \cite{Pushkarev2009}, respectively.  Figure\,\ref{error_analysis} shows the two dimensional error contours for the shear\,(iii) model.  Since our shear models produce better fits with high p-values and a reasonable value of $a$, we conclude that it is likely that relativistic shear and Doppler beaming play a role in jet appearance.  As our final result for a measurement of $\Gamma\theta_j$, we report $0.21 \pm 0.03$ from the shear\,(iii) model.  However, for a more direct comparison between our value of $\rho$ and that of other researchers, we calculate the expected value of $\rho_{\rm eff}(\tapp)$
\begin{align}
\langle \rho_{\rm eff} \rangle &= \int \rho_{\rm eff}(\tapp)P(\tapp)d\tapp \notag \\
& \approx 0.13\pm 0.02,
\label{ave_rho}
\end{align}
where the parameters of $P(\tapp)$ are those for the shear iii model listed in Table\,\ref{bestfits}, and the confidence interval comes from the bootstrap simulations used to derive the confidence intervals for the shear iii model.  Indeed, this value of $\langle \rho_{\rm eff}\rangle \approx 0.13 \pm 0.02$ is consistent with both \cite{Pushkarev2009} and \cite{Jorstad2005}.

A more rigorous comparison between our best-fit value for $\rho$ and those of other researchers requires a proper error analysis for all of the different measured/inferred values for $\rho$, both in this paper and in other works.  However, in the case of \cite{Jorstad2005} and \cite{Pushkarev2009}, the error in their estimated values for $\theta_j$ and $\Gamma$ for each jet is unknown, as these estimates rely upon highly uncertain model assumptions regarding, for example, equipartition brightness temperature arguments and the equation of component variability with light-crossing times.  In addition to this error, it is also possible that AGN jets possess a range of values of $\rho$, as opposed to our assumption that all jets have the same $\rho$.  These same issues apply to our simple model.  More specifically, our confidence limits (see Fig.\,\ref{error_analysis}) are probably underestimated since there is considerable uncertainty in our model of velocity shear, our assumption of jet conical geometry, and our assumption that all MOJAVE jets posses the same value of $\rho$.

\section{Discussion}
\label{discussion}
\subsection{Relativistic jet physics and \mbox{\boldmath{$\Gamma\theta_{\lowercase{j}}$}}}
\label{jetphysics}
A variety of physical processes in jets are sensitive to $\Gamma\theta_j$.  Causal connection implies that $\theta_j\mathcal{M}\leq 1$, where $\mathcal{M}=\beta\Gamma/(\Gamma_s\beta_s)$ is the relativistic Mach number, which is the ratio of the jet proper speed to proper signal speed.  (In the Appendix we explain the relationship between $\theta_j\mathcal{M}$, $\Gamma\theta_j$, and causality).  For jets with a dynamically important magnetic field, the signal speed is the fast magnetosonic speed, while for jets with no significant magnetic field the signal speed is the sound speed, which is $\beta_s=1/\sqrt{3}$ for a relativistically hot jet.  The fast magnetosonic proper speed is $\Gamma_{ms}\beta_{ms}=\sigma^{1/2}$, where $\sigma$ is the magnetization parameter, which is the ratio of Poynting flux to kinetic flux \citep{Kennel1984}.  Beyond the acceleration zone the jet is likely to be in equipartition such that $\sigma$ is of the order of unity \citep[e.g.,][]{Komissarov2007}.  Thus, fiducial jet signal speeds imply that the causal connection condition for jets is
\begin{align}
\Gamma\theta_j &\la 0.7 \quad &&\mbox{relativistic sound speed} \notag \\
\Gamma\theta_j &\la 1 \quad &&\mbox{equipartition fast magnetosonic speed}, \notag
\end{align}
indicating that AGN jets with the value we have determined here, $\Gamma\theta_j\sim 0.2$, are probably causally connected.  This verifies the standard picture of jet production \citep[see Sect.\,\ref{grb} and ][]{Komissarov2009}, and also implies that AGN jets are susceptible to various instabilities and reconnection, and are also sensitive to conditions at the boundary between the jet and the interstellar medium.

The particular value of $\Gamma\theta_j$ is also important for instability development \citep{Narayan2009}, because it gives a measure of the extent to which jet sidewise expansion inhibits instability growth.  Most global instabilities grow on some comoving signal crossing timescale, $t_{\rm dyn}=\Gamma\theta_j z/(\beta_sc)$, where $\beta_s$ is the signal speed and $z$ is the jet height above the launching region. For Kelvin-Helmholtz instabilities $\beta_s$ is the sound speed \citep{Perucho2004,Hardee2005}, while for current driven kink instabilities $\beta_s$ is the \alfven{} speed \citep{Giannios2006}.  This so-called dynamical timescale for the growth of instabilities must be compared to the jet expansion time $t_{\rm exp}=z/(\beta c)$, where if $t_{\rm exp}>t_{\rm dyn}$ then jet expansion will quench instability growth \citep{Begelman1998,Giannios2006,Moll2008,Spruit2010}.  This criterion for instability growth then becomes
\begin{align}
\frac{t_{\rm dyn}}{t_{\rm exp}}\approx\frac{\Gamma\theta_j}{\beta_s}\lesssim1.
\end{align}
Since $\Gamma\theta_j\sim 0.2$, then AGN jet instabilities can grow despite jet expansion, unless $\beta_s\lesssim0.2$, which would be below the relativistic adiabatic sound speed of $\sim0.6$.  

\subsection{Jet acceleration for GRBs vs.~AGNs}
\label{grb}
Long duration gamma-ray bursts (GRBs) have typical values of $\Gamma\theta_j$ of the order of $10-30$ \citep{Panaitescu2002}, which is two orders of magnitude higher than the value we find for AGN jets:
\begin{align}
\frac{(\Gamma\theta_j)_{\rm GRB}}{(\Gamma\theta_j)_{\rm AGN}}\sim 100. \notag
\end{align}
Thus, it is possible that different physics are at work in GRB jets.  \cite{Tchekhovskoy2010} found that a jet acceleration mechanism (first discovered by \cite{Aloy2006} and \cite{Mizuno2008}) can operate in GRBs in which a brief period of bulk acceleration occurs upon jet break out into the circumstellar medium and allows the jet to take on values of $\Gamma\theta_j\gg 1$.  \cite{Komissarov2010} call this process {\it rarefaction acceleration}, and contrast it to the standard jet acceleration model of {\it collimation acceleration} described in \cite{Li1992} and many other works.  Collimation acceleration entails jet acceleration over an extended distance along the jet \citep{Vlahakis2004} and implies that $\Gamma\theta_j\leq 1$ \citep{Komissarov2009,Tchekhovskoy2009}.  In contrast, rarefaction acceleration occurs in GRBs because they are initially confined by the shocked boundary layer in the star until the jet breaks out into the circumstellar medium, becomes unconfined, and launches a rarefaction wave toward the center of the jet.  If the jet is still magnetically dominated at that point, then this process further accelerates the jet, and can produce $\Gamma\theta_j\gg 1$.  Thus, the dichotomy between GRB jets and AGN jets is nicely explained by the different physical processes at work in GRB jets (rarefaction acceleration due to jet break out) and AGN jets (collimation acceleration).  Furthermore, as \cite{Komissarov2010} explain, rarefaction acceleration increases the $\Gamma\theta_j$ parameter primarily by increasing $\Gamma$ alone (The increase in $\theta_j$ is less than $1/\Gamma$).  Thus, to the degree that the typical AGN jet value of $\Gamma\theta_j$ is equal to the pre-breakout GRB jet value of $\Gamma\theta_j$, one can infer that $(\Gamma\theta_j)_{\rm GRB}$ is so much larger than $(\Gamma\theta_j)_{\rm AGN}$ because of the increase of the GRB jet's $\Gamma$ during the rarefaction acceleration process.  Interestingly, this suggests that GRBs have $\Gamma\gtrsim 100$, which is consistent with the lower limit obtained by requiring that GRB prompt emission regions be optically thin to gamma-rays with respect to photon-photon pair production \citep[e.g.,][]{Piran2004}.

\subsection{Doppler beaming and \mbox{\boldmath{$\tapp$}}}
\label{dopplerbeaming}
If $\Gamma\theta_j$ is approximately constant in the AGN jet population, then $\tapp$ is an important observable quantity related to Doppler beaming for two reasons.  First, it can serve as a dividing line between highly beamed ($\theta_{ob}<1/\Gamma$) jets and not highly beamed jets ($\theta_{ob}>1/\Gamma$), a property we exploit in our model of velocity shear in Sect.\,\ref{shear}.  This division conveniently maps onto $\tapp$ as follows:
\begin{align}
\tapp &>\arctan(\rho) \Longleftrightarrow \mbox{highly beamed jets} \notag \\
\tapp&<\arctan(\rho) \Longleftrightarrow \mbox{not highly beamed jets}. \notag
\end{align}
This implies that jets with $\tapp\sim \Gamma\theta_j\sim 0.2$ are observed at the critical angle, thus maximizing the apparent speed of their superluminal components.  This division provides a concise way of explaining the \cite{Pushkarev2009} argument that large-opening angle jets are more highly Doppler beamed: large opening angle jets with $\tapp \gtrsim 0.2$ are all observed within the critical angle and therefore more highly beamed than smaller $\tapp$ jets.  \cite{Pushkarev2009} also find that AGN jets with larger $\tapp$ have a higher \emph{Fermi}-LAT detection rate, and all jets with $\tapp>0.35$ are detected by \emph{Fermi}-LAT, implying that \emph{Fermi}-LAT detected jets tend to have higher Doppler factors.  Notably, this finding is also supported by \cite{Kovalev2009}, \cite{Savolainen2010}, and \cite{Lister2009c}, who found evidence that radio jets of Fermi-detected AGN are more likely to have high Doppler factors than are non \emph{Fermi}-LAT detected sources. 

Second, by measuring both a jet's $\tapp$ and its typical apparent speed $\beta_{\rm app}=\beta\sin{\theta_{ob}}(1-\beta\cos{\theta_{ob}})^{-1}$, we can derive that jet's Doppler factor, Lorentz factor, and therefore also the viewing angle,
\begin{align}
&\delta=\frac{\beta_{\rm app}\tan{\tapp}}{\beta\rho_{\rm eff}}\approx\frac{\beta_{\rm app}\tan{\tapp}}{\rho_{\rm eff}} \label{delta}\\
&\Gamma\approx\frac{\beta_{\rm app}\left(1+\rho_{\rm eff}^2\cot^2{\tapp}\right)}{2\rho_{\rm eff}\cot{\tapp}} \label{Gamma}\\
&\theta_{ob}\approx\frac{2\rho_{\rm eff}^2\cot^2{\tapp}}{\beta_{\rm app}(1+\rho_{\rm eff}^2\cot^2{\tapp})} \label{theta_ob},
\end{align}
where $\delta=(\Gamma-\sqrt{\Gamma^2-1}\cos{\theta_{ob}})^{-1}$ is the Doppler factor and $\beta$ is the jet velocity in units of the speed of light.  Except for Eq.\,(\ref{delta}), which shows both the exact and approximate form of $\delta$, the above equations are approximations that assume $\Gamma\gg1$ and $\theta_{ob}\ll 1$.   

Thus, Eqs.\,(\ref{delta}--\ref{theta_ob}) demonstrate that, if the spread of $\rho$ ($=\Gamma\theta_j$) is small enough in the jet population, the measurable quantities $\beta_{\rm app}$ and $\tapp$ can be useful for calculating intrinsic jet quantities such as the intrinsic brightness temperature and Lorentz factor of jet components.  In a future work we intend to explore this new method of deriving a jet's beaming parameters.

\section{Conclusion}
\label{conclusion}
We have derived a statistical model of relativistic jet apparent opening angles and fit it to the observed distribution of jet apparent opening angles in the MOJAVE sample.  The product of Lorentz factor and intrinsic jet opening angle $\Gamma\theta_j$ is a free parameter in our model and was determined by the best fit to be $\Gamma\theta_j\approx 0.2$.  We summarize our conclusions as follows.
\begin{enumerate}

\item $\Gamma\theta_j\sim 0.2$ implies that jets are causally connected (see the Appendix), which is predicted by magnetic jet production models.  Causal connection also implies that AGN jets are subject to Kelvin-Helmholtz and current-driven (kink) modes, unless the relevant signal speed is $\la \Gamma\theta_jc \sim 0.2c$. \\

\item The value of $\Gamma\theta_j$ for GRB jets is 100 times larger than $\Gamma\theta_j$ for AGN jets.  This difference is neatly explained by an acceleration process probably unique to GRBs, wherein a rarefaction wave is launched into the jet after the jet breaks out of its stellar envelope and into the lower pressure circumstellar medium.  This is consistent with the high Lorentz factors inferred for GRB jets of $\Gamma \gtrsim 100$.\\

\item In order to adequately fit the $\tapp$ data, we included the effects of relativistic velocity shear and Doppler beaming.  Velocity shear affects jets by making blazars appear narrower as their ultra-relativistic inner spine is all that is visible, while jets viewed outside the critical angle $1/\Gamma$ appear to have larger jet opening angles.  Distinguishing jets based on their critical angle conveniently creates a division between highly beamed jets and not so highly beamed jets that corresponds to whether an individual jet's apparent opening angles is $\tapp \lesssim \Gamma\theta_j$ (not highly beamed) or $\tapp\gtrsim \Gamma\theta_j$ (highly beamed).\\

\item Assuming $\Gamma\theta_j$ is mostly constant across the AGN jet population, then a jet's Doppler factor, Lorentz factor, and viewing angle can be calculated if the observable values of apparent jet opening angle $\tapp$ and the apparent speed of the jet components $\beta_{\rm app}$ are known.  This is shown in Eqs.\,(\ref{delta}--\ref{theta_ob}).

\end{enumerate}
\section*{Acknowledgements}
ECB thanks M.~B\"{o}ck, M.~Zamaninasab, M.~Lister, M.~Lyutikov, and D.~Giannios for valuable discussions.  ABP was supported by the ``Non-stationary processes in the Universe'' Program of the Presidium of the Russian Academy of Sciences.  YYK was supported by the Russian Foundation for Basic Research (project 12-02-33101), the Dynasty Foundation, and the Research Program OFN-17 of the Division of Physics, Russian Academy of Sciences.  This research has made use of data from the MOJAVE database that is maintained by the MOJAVE team \citep{Lister2009}.

\appendix
\section{Causal connection and \mbox{\boldmath{$\Gamma\theta_{\lowercase{j}}$}}}
\label{causality}
Here we discuss two different criteria for defining whether or a not a jet is causally connected for the simplistic case of a jet with radial velocity streamlines of constant speed.  For jets with velocity shear, as we posit in this work, causal connection is more complicated than we have presented below, although we expect the following discussion to be approximately correct. First, the relativistic Mach number $\mathcal{M}=\beta\Gamma/(\beta_s\Gamma_s)$ \citep{Konigl1980} is sometimes used to define causal connection for supersonic jets by requiring $\theta_j\mathcal{M}<1$, or $\Gamma\theta_j<\Gamma_s\beta_s/\beta$ \citep[e.g.,][]{Komissarov2009}.  Second, causally connected jets are sometimes defined as those for which $\Gamma\theta_j<1$ \citep[e.g.,][]{Zakamska2008}.  We note that for relativistic jets with $\beta\approx 1$ and an ultra-relativistic equation of state where the proper sound speed is $\Gamma_s\beta_s=2^{-1/2}\approx 0.71$, both of these causality criteria resemble one another: $\Gamma\theta_j<0.71$ for the relativistic Mach number approach, and $\Gamma\theta_j<1$ for the other.  However, for magnetically dominated plasmas, the fast magnetosonic speed can approach the speed of light, thus no upper bound can be placed on $\Gamma_s\beta_s$ and these two criteria for causal connection give contradictory answers.  Thus, a jet can have any value of $\Gamma\theta_j$ and still in principle be causally connected, provided it is magnetically dominated enough.  In particular, for jets with magnetization $\sigma$ \citep{Kennel1984}, the proper \alfven{} speed is $\Gamma_{\rm A}\beta_{\rm A}=\sigma^{1/2}$, thus the Mach number condition for the jet to be causally connected is $\Gamma\theta_j<\sigma^{1/2}$ \citep[e.g.,][]{Komissarov2009}.  Jets can then be causally connected even though $\Gamma\theta_j\gg1$, as long as $\sigma$ is large enough (high $\sigma$ implies the jet is Poynting flux dominated).  Below, we discuss why these two criteria are different and conclude that the relativistic Mach angle analysis is usually the more appropriate criterion, even though it is only approximate.\\

\noindent
\textbf{\mbox{\boldmath{$\theta_{\lowercase{j}}\mathcal{M}<1$}} criterion:}\\
This criterion is only relevant for highly supersonic or supermagnetosonic jets, since for subsonic or transonic jets there is no limit on wave propagation.

The relativistic Mach number can be derived by assuming a flow with parallel velocity streamlines and by analyzing the observer frame angle a sound wave can make with respect to the flow direction, $\tan\chi=\beta_{\perp}/\beta_{\|}$.  The relativistic Mach angle is then found by maximizing $\chi$ by varying $\chi'$, the rest frame angle between the flow direction and the sound wave direction.  This procedure gives $\cos\chi'=-\beta_s/\beta$ and a maximum angle of $\sin\chi_{\rm max}=1/\mathcal{M}$ \citep{Konigl1980}.  Thus, $\chi_{\rm max}$ represents the largest observer frame angle a sound wave can make with respect to a supersonic flow.  For this reason, jets where $\theta_j>\chi_{\rm max}$ are assumed to be out of causal contact with themselves.  However, this Mach angle analysis assumes a flow of parallel velocity streamlines, something that is not the case for conical jets.

For conical jets, jet sidewise expansion lengthens the signal crossing time compared to a flow with parallel streamlines (e.g., a cylindrical jet).  Causal connection can be analyzed by making a simple estimate of a conical jet's signal crossing time, assuming the signal propagates at a speed $\beta_s$ in the local fluid rest frame and makes an angle $\chi'$ between the jet comoving frame signal wave propagation direction and the local fluid streamline. We also assume that the jet flow is radial with a half-opening angle of $\theta_j$ and a constant Lorentz factor of $\Gamma$.  In the observer frame the wave has a speed parallel to the local streamline of 
\begin{align}
\beta_{r}=\frac{\beta_s\cos\chi'+\beta}{1+\beta\beta_s\cos\chi'}
\end{align}
and a perpendicular speed of
\begin{align}
\beta_{\theta}=\frac{\beta_s\sin\chi'}{\Gamma(1+\beta\beta_s\cos\chi')}.
\end{align}
For simplicity we assume the emitted wave trajectory is such that $\chi'$ is constant.  If the radial coordinate (centered on the jet's central engine) of the wave is $r$, then $dr=\beta_{r}c dt$.  In a time $dt$, the wave propagates in the polar direction an arc length of $ds=\beta_{\theta}c dt$.  In terms of polar angle, then, the signal propagation can be written as $d\theta=\beta_{\theta}cdt/r$, which, combined with the $dr=\beta_{r}c dt$, can be solved for the time it takes for a signal to propagate through a polar angle $\theta_j$,
\begin{align}
t_{\rm cross}=\frac{r_0}{\beta_r c}\left(\exp\left(\frac{\theta_j\beta_r}{\beta_{\theta}}\right)-1\right),
\label{cross}
\end{align}
where $r_0$ is the radial location of the initial wave emission.  We note that a different form of Eq.\,(\ref{cross}) was derived in \cite{Kinoshita2004} and a similar result was also found in \cite{Nakar2003}.  If in the observer frame a sound wave has a trajectory such that $\chi=1/\mathcal{M}$ (i.e., $\chi'=-\beta_s/\beta$), then $\beta_r/\beta_{\theta}\approx \mathcal{M}$ and $\beta_r\approx \beta$, yielding the differential equation $d\theta=(\mathcal{M} r)^{-1}dr$.  This differential equation has the solution for polar angle through which the signal propagates of $\theta(r)=\mathcal{M}^{-1}\ln(r/r_0)$ with the associated signal crossing time of
\begin{align}
t_{\rm cross}&\approx\frac{r_0}{\beta c}\left(e^{\theta_j \mathcal{M}}-1\right),
\label{tdyn}
\end{align}
assuming $\mathcal{M}\gg1$.  Eq.\,(\ref{tdyn}) implies that jets for which $\theta_j\mathcal{M}\gg1$ have $t_{\rm cross}\approx (r_0/c) \exp (\theta_j\mathcal{M})$, effectively making such jets fall out of causal contact in the sense that the dynamic time is longer than the jet expansion time $r_0/ c$ by a factor $ \exp (\theta_j\mathcal{M})$.  

Accelerating conical jets are different in that they can have a causal horizon that depends on the details of jet acceleration \citep{Kinoshita2004}.  An accelerating supersonic jet will have an increasing proper speed and a decreasing (or constant) proper signal speed, which we parameterize as $\mathcal{M}=\mathcal{M}_0(r/r_0)^b$.  For a signal emitted at $r_0$ that propagates at the Mach angle relative to the local fluid streamline, then $d\theta=(\mathcal{M} r)^{-1}dr$ has the asymptotic solution
\begin{align}
\theta_{\infty}=\frac{1}{b\mathcal{M}_0}.
\label{accel}
\end{align}
That is, disturbances located at radius $r_0$ will propagate through a polar angle $\theta_{\infty}$ as $r\rightarrow \infty$. Thus, in this circumstance, the causal connection criterion becomes $\theta_j\mathcal{M}(r)<b^{-1}$.  \\

\noindent
\textbf{\mbox{\boldmath{$\Gamma\theta_{\lowercase{j}}<1$}} criterion:}\\
We assume an initially cylindrical jet with speed $\beta\approx 1$ and associated Lorentz factor $\Gamma\gg 1$, and let the radius of the cylindrical flow suddenly begin to expand in the flow rest frame with velocity $\beta_{\rm exp}'$ perpendicular to the symmetry axis.  Transforming back into the observer frame then gives $\beta_{\perp}=\beta_{\rm exp}'/\Gamma$, the small angle the velocity stream lines make with the jet axis is now \begin{align}
\theta=\frac{\beta_{\rm exp}'}{\Gamma}
\end{align}
The requirement that the jet cross section expands at less than the speed of light $\beta_{\rm exp}'<1$ implies that \citep{Zakamska2008}
\begin{align}
\Gamma\theta_j<1.
\end{align}
Thus, requiring that $\Gamma\theta_j<1$ is an important constraint for jet flows that are initially close to cylindrical, and for some reason undergo expansion.  We note, however, that this constraint is not important for flows that are initially not collimated, such as the highly relativistic equatorial outflows from pulsars that power pulsar wind nebulae \citep{Kennel1984}.

Alternatively, $\Gamma\theta_j$ can be an important quantity if the sound wave emission direction in the local rest frame is defined as perpendicular to the local flow direction, i.e., $\chi'=\pi/2$, as could be the case if the wave is restricted to a thin spherical shell \citep[which may be relevant for GRBs,][]{Lyutikov2003,Kinoshita2004}, so that $\beta_{\theta}=\beta_s/\Gamma$ and $\beta_r=\beta$.  In this case, the sound crossing time from Eq.\,(\ref{cross}) becomes
\begin{align}
t_{\rm cross}=\frac{r_0}{\beta c}\left(\exp\left(\frac{\Gamma\theta_j\beta}{\beta_s}\right)-1\right).
\label{tdyn2}
\end{align}
For jets with $\beta\approx 1$, $\beta_s\approx 1$, and $\Gamma\theta_j \gg 1$, then $t_{\rm cross}\approx (r_0/c) \exp (\Gamma\theta_j)$, showing that in this case $\Gamma\theta_j$ plays the same role that $\theta_j\mathcal{M}$ does in the general case for determining an effective causal condition.

\label{lastpage}

\end{document}